\documentstyle[prd,aps]{revtex}

\begin{document} \title{Chaos and a Resonance Mechanism for Structure
Formation in Inflationary Models\footnote{This essay received an
"honorable mention" from the Gravity Research Foundation, 1998 - Ed.}}

\author{H. P. de Oliveira\thanks{E-mail: oliveira@symbcomp.uerj.br}}

\address{{\it Universidade do Estado do Rio de Janeiro }\\ {\it Instituto
de F\'{\i}sica - Departamento de F\'{\i}sica Te\'orica,}\\ {\it CEP
20550-013 Rio de Janeiro, RJ, Brazil}}

\author{I. Dami\~ao Soares\thanks{E-mail: ivano@lca1.drp.cbpf.br}}

\address{{\it Centro Brasileiro de Pesquisas F\'\i sicas\\ Rua Dr.
Xavier Sigaud, 150 \\ CEP 22290, Rio de Janeiro~--~RJ, Brazil}}

\maketitle

\thispagestyle{empty}

\label{address} \vspace{0.34cm} \noindent

\begin{abstract} We exhibit a resonance mechanism of amplification of
density perturbations in inflationary mo-dels, using a minimal set
of ingredients (an effective cosmological constant, a scalar field
minimally coupled to the gravitational field and matter), common to
most models in the literature of inflation. This mechanism is based
on the structure of homoclinic cylinders, emanating from an unstable
periodic orbit in the neighborhood of a saddle-center critical point,
present in the phase space of the model. The cylindrical structure
induces oscillatory motions of the scales of the universe whenever the
orbit visits the neighborhood of the saddle-center, before the universe
enters a period of exponential expansion. The oscillations of the scale
functions produce, by a resonance mechanism, the amplification of a
selected wave number spectrum of density perturbations, and can explain
the hierarchy of scales observed in the actual universe. The transversal
crossings of the homoclinic cylinders induce chaos in the dynamics of the
model, a fact intimately connected to the resonance mechanism occuring
immediately before the exit to inflation.  \end{abstract}

\label{PACS number:47.75+f}

\vfill\eject

\newpage

The existence of an inflationary phase in the early stages of our
Universe has become one of the paradigms of modern cosmology, and is
now being subject to experimental verification through the crucial
measurements of small scale anisotropies in the cosmic background
radiation. The basic physical ingredient for this inflationary phase
is the existence of a scalar field - the inflaton field, the vacuum
energy of which plays the role of a cosmological constant, engendering
via the gravitational dynamics an exponential expansion in the comoving
scales of the universe. This phase is considered to have evolved from a
pre-inflationary phase just exiting the Planck era, the matter content
of which may be thought to be modeled by radiation plus a scalar
field minimally coupled to the gravitational field. To be considered
successful, the inflationary paradigm must accomodate models in which
initial homogeneities are generated,  leading to the formation of the
structure present in our actual universe. In the literature of inflation,
it has been claimed that small fluctuations in the scalar field may give
rise to an almost scale-free spectrum of adiabatic density perturbations
(the Harrison-Zeldovich spectrum), which is the standard spectrum used in
explaining galaxy formation. However the number of distinct large scale
objects and structures in our universe, as galaxies, cluster of galaxies,
quasars, filaments and walls, is very large and the sizes of these objects
form a hierarchy of scales that is not described by the flat spectrum of
the initial perturbations\cite{linde}. In other words, the scale-free
adiabatic perturbations appears not to be sufficient to produce the
large-scale structure of the observed universe. We are therefore led to
pose the following basic question: would it be possible, in the realm
of the inflationary paradigm, to obtain adiabatic  perturbations with a
spectrum having several maxima, which would help to  explain the origin
of the hierarchy of scales?

Our purpose in this essay is to give a positive answer to the above
question, and show that the ingredients present in most inflationary
models may engender a resonance mechanism, in the beginning of,
or immediately before the inflationary period, that amplifies scalar
field perturbations with a selected wave number spectrum. We emphasize
the simplicity of this mechanism, which deals with a minimum set
of ingredients only and does not demand the introduction of further
fields, fine tuning of parameters and special initial conditions. Even
if the universe inflates afterwards, the relative rate of amplitudes
produced after the resonance amplification would be maintained as an
imprint in the initial spectrum of density fluctuations. We will also
show that this resonance amplification mechanism has a fundamental
connection to the chaotic behaviour of the model, and its chaotic exit
to inflation\cite{oss}. Similar resonance mechanisms have been previously 
applied for particle production during reheating in the realm of inflationary
models\cite{brand}.  

Let us consider the following configuration, that may constitute a
simple description of a preinflationary phase of the universe in most
of the inflationary models discussed in the literature. We start from a
Bianchi IX model characterized by two scale factors $A(t)$ and $B(t)$,
with a perfect fluid and a cosmological constant $\Lambda$ playing
the role of the vacuum energy of the scalar field. For simplicity, we
consider the case of dust, but the basic features are not altered for
other types of perfect fluid like radiation, for instance. According
with Ref. \cite{oss}, the full dynamics of the models is governed by
the Hamiltonian

\begin{equation} H(A,B,P_A,P_B ) = \frac{P_A\,P_B}{4\,B} -
\frac{A\,P_A^2}{8\,B^2} + 2\,A - \frac{A^3}{2\,B^2} - 2\,\Lambda\,A\,B^2 -
E_0 = 0, \end{equation}

\noindent where $P_A$ and $P_B$ are the momenta canonically conjugated
to $A$ and $B$, respectively, and $E_0$ is a constant proportional to the
total matter energy of the models. The associated dynamical system has one
critical point $E$ of saddle-center type in the finite region of the phase
space. The phase space has two critical points at infinity, corresponding
to the stable and unstable de Sitter solutions, and the scale factors
$A(t)$ and $B(t)$ approach the de Sitter attractor exponentially, $A =
B \cong e^{\sqrt{\Lambda/3}\,t}$. The dynamical system generated by (1)
admits the invariant manifold ${\cal{M}}$ ($A =B$, $P_A = \frac{P_B}{2}$)
in which all orbits are Friedmann models with cosmological constant and
dust. The critical point $E$ is contained in ${\cal{M}}$, and that the
separatrices $S$ constitute boundaries between isotropic models that
collapse or escape to the de Sitter configuration. The models to be
considered in this essay are the ones restricted to a neighborhood of
the invariant manifold ${\cal{M}}$ and can be understood as describing
the early stages of inflation in which anisotropy is present, in the
form of small fluctuations.

The basic point of the amplification mechanism to be discussed here is
the structure of homoclinc cylinders present in the neighborhood of the
saddle-center. To see this we make use of a theorem by Moser\cite{moser}
which states that it is always possible to find a set of canonical
variables, say $(q_1, q_2, p_1, p_2)$, such that, in a small neighborhood
of a saddle-center at the origin, the Hamiltonian is separable as

\begin{equation}
H(q_1,q_2,p_1,p_2)=\frac{\sqrt{\Lambda}}{2}\,(p_2^2-q_2^2) -
\sqrt{2\,\Lambda}\,(p_1^2+q_1^2)  - E_0 + E_{cr} \approx 0, \end{equation}

\noindent where $E_{cr}$ is the energy associated to the saddle-center,
with the two approximatte constants of motion given by the partial
energies

\begin{equation} E_{hyp}=\frac{\sqrt{\Lambda}}{2}\,(p_2^2-q_2^2), \;
E_{rot} = \sqrt{2\,\Lambda}\,(p_1^2+q_1^2).  \end{equation}

\noindent The energies $E_{hyp}$ and $E_{rot}$ will be referred to
as the hyperbolic motion  energy and rotational motion energy of the
system about $E$, respectively. The rotational motion corresponds to
periodic orbits in a linear neighborhood of $E$. For zero energy of
the hyperbolic motion, we have the linear stable $V_s$ and unstable
$V_u$ one-dimensional manifolds. The separatrices $S$ in the invariant
manifol ${\cal{M}}$ are actually the non-linear extension of $V_s$
and $V_u$. The direct product of a periodic orbit with $V_s$ and $V_u$
generates, in the linear neighborhood of $E$, the structure of stable
and unstable cylinders, which coalesce into the periodic orbit for large
positive and negative times, respectively\cite{ozorio}. A general orbit 
which visits the neighborhood of $E$ has an oscillatory approach to the 
cylinders, the closer as the energy of the hyperbolic motion goes to zero. 
Away from the linear neighborhood of $E$, the linear approximation is no
longer valid. Higher order terms become important for the dynamics,
and the non-integrability of the system results in the distortion and
twisting of the cylinders. The stable cylinder and the unstable one will
cross each other transversally, producing chaotic sets\cite{oss} in the
neighborhood of the invariant manifold ${\cal{M}}$; orbits with initial
conditions taken on these chaotic sets will be highly sensitive to small
perturbations of the initial conditions. The chaotic character manifests
itself physically such that a small perturbation of initial conditions
taken on these sets would change an orbit from collapse to escape into
the de Sitter phase, and vice-versa. Also, it is always possible to
select initial conditions in these sets such that the scale factors
oscillate an arbitrary fixed time before escaping to a de Sitter phase.

The physical effects related to the growing of inhomogeneous perturbations
occur during the oscillatory phase when a given orbit approach $E$
before to collapse or to escape. We consider a given orbit initially in
a very small sphere of initial conditions about the invariant manifold
${\cal{M}}$. In this regime the scale factors oscillate in a small region
about the critical point $E$. To avoid technical dificulties concerning
the perturbative formalism in anisotropic backgrounds, and motivated
by the fact that in the overall dynamics of perturbations these small
anisotropy fluctuations can be averaged, we introduce the average scale
factor $l(t)$ of the universe defined by

\begin{equation} l = (A\,B^2)^{1/3}.  \end{equation}

\noindent This approximately corresponds to a closed
Friedamnn-Robertson-Walker backgound with $l(t)$ as the scale factor,
either in the oscillatory regime and in the inflationary phase. In a
very small neighborhood of $E$ the approximate analytical solution for
the average scale factor is

\begin{equation} l(t) \approx l_0 + C_1\,e^{t/2\,l_0} + C_2\,e^{-t/2\,l_0}
+ \epsilon\,cos\,(2\,\omega\,t) \label{sf} \end{equation}

\noindent where $l_0 = 1/\sqrt{4\,\Lambda}$, $\omega = \sqrt{2\,\Lambda}$,
and $\epsilon$, $C_1$ and $C_2$ are very small arbitrary constants of the
same order of $R$. The exponential terms are associated to the hyperbolic
energy mode, while the last term is associated to the rotational energy
mode. The pure oscillatory phase is characterized by $C_1 = C_2 = 0$
(on the cylinder), which is a very special class of solution, hardly
realized by physically reasonable universes. The escape to inflation
and the collapse depend on the relative signs of $C_1$ and $C_2$, but
the smaller are these constants the larger is the time that the orbit
spends in the oscillatory phase. Nevertheless it must be stressed that in
this oscillating phase the universe has periods of accelerated expansion
followed by accelerated contraction, and the escape to inflation actually
corresponds to a progressive increase of the inflationary phase.

The next step is to study the behavior of inhomogeneous perturbations
of matter $\delta\,\rho$ and scalar field $\delta\,\varphi$ in this
background. Before presenting the evolution equations for the perturbed
quantities, it is important to stress that $\delta\,\varphi$ will
give rise to second order perturbations of the energy of the scalar
field, $\delta\,\rho_{\varphi}$, since $\dot{\varphi}^{(0)} \approx
V^{\prime}(\varphi^{(0)}) \approx 0$, with $\varphi^{(0)} \approx
constant$ being the scalar field of the background corresponding to an
extremum of the potential. As a consequence, there will be no coupling
between $\delta\,\rho$ and $\delta\,\rho_{\varphi}$, once the former
represents first order deviations.

Introducing the contrast density $\chi = \frac{\delta\,\rho}{\rho}$,
one can show that its evolution equation is given by

\begin{equation} \ddot{\chi_k} + 2\,\frac{\dot{l}}{l}\,\dot{\chi_k} -
\frac{1}{2}\,\rho\,\chi_k = 0. \label{evol1} \end{equation}

\noindent Here $\chi_k$ is the mode associated to
the scalar harmonic basis $Q_{(k)}(\bf{x})$ which
satisfies $h^{\mu\nu}\,Q_{(k)}({\bf{x}})_{,\mu;\nu} =
\frac{k^2}{l^2}\,Q_{(k)}(\bf{x})$, $h^{\mu\nu}$ being the projector into
the spatial surface, and $k^2 = n^2 -1$ with $n \geq 1$ assuming integers
values.  $\rho = E_0/l(t)^3$ is the energy density of the background. In
the oscillatory phase about $E$ (Eq. (\ref{sf}) with $C_1 = C_2 = 0$),
the above equation, after a proper rescaling of the variable $\chi_k$,
assumes the form

\begin{equation} \ddot{\chi_k} + (-q + a\,cos\,(2\,\omega\,t))\,\chi_k =
0, \end{equation}

\noindent where $q$ is a positive constant, and $|a| << q$. This is
the Mathieu equation inside the parametric region of instability,
in which all modes $k$ are equally amplified. If the universe, after
this oscillating phase, undergoes to a successful inflation, $l \approx
exp\,{\sqrt{\Lambda/3}\,t}$, and the equation for all modes $\chi_k$
reduces to

\begin{equation} \ddot{\chi_k} + \frac{2}{3}\,\sqrt{\Lambda}\,\dot{\chi_k}
\approx 0.  \end{equation}

\noindent As a consequence all modes $\chi_k$ will decrease
asymptotically, or in other words, all perturbations of matter produced
during the oscillatory phase will be washed out in the inflationary
regime, and cannot therefore be considered as seeds of structure
formation.

Now we consider the evolution of perturbations of the
scalar field $\varphi({\bf{x}},t) = \varphi^{(0)}(t) +
Q_{(k)}({\bf{x}})\,\delta\,\varphi_{k}(t)$, where the background
$\varphi^{(0)}(t) \approx constant$. The equation of motion for the
perturbed modes of the scalar field is

\begin{equation} (\delta\,\varphi_{k}\ddot{)} +
3\,\frac{\dot{l}}{l}\,(\delta\,\varphi_{k}\dot{)}
+ \left(V^{\prime\prime}(\varphi^{(0)}) +
\frac{k^2}{l^2}\right)\,\delta\,\varphi_{k} = 0. \label{eq1}
\end{equation}

\noindent Introducing the variable $\delta_k = l^{3/2}\,\delta\,\varphi_k$
into the above equation as well as the approximate solution in the
oscillatory regime, it follows

\begin{equation} \ddot{\delta_{k}} + \left[
\left(V^{\prime\prime}(\varphi_0) + \frac{k^2}{l_0^2}
\right) - \epsilon\,\left(\frac{2\,k^2}{l_0^4}
+\frac{6\,\omega^2}{l_0^2}\right)\,cos\,(2\,\omega\,t)\right]\,\delta_{k}
= 0, \end{equation}

\noindent that is a typical Mathieu equation. As an important property
of solutions of this equation is the existence of an instability zone
in the parameter space in which the solutions grow exponentially as
$\delta_k exp\,(\nu_k\,t)$. This instability is known by {\it parametric
resonance}. For small $\epsilon$, the condition for parametric resonance
is given by

\begin{equation} \frac{V^{\prime\prime}(\varphi^{(0)})}{2\,\Lambda} +
2\,k^2 = j^2 + \Delta(\epsilon)  \label{max} \end{equation}

\noindent where $j$ is an arbitrary integer and $\Delta(\epsilon)$ is a
small adimensional number dependent of $\epsilon$. This means that only
the modes with $k$ satisfying the above relation grow exponentially. These
modes are determined by the ratio between the mass of the scalar field
$V^{\prime\prime}(\varphi^{(0)})$, and the vacuum energy of the scalar
field. The ratio $\frac{V^{\prime\prime}(\varphi^{(0)})}{2\,\Lambda}$
is a fundamental parameter in inflationary models; since the resonance
mechanism is believed to be general, the order of this parameter could
be evaluated through Eq. (\ref{max}) once the spectrum is experimentally
determined.  It will be quite ilustrative to choose standard values used
for the afore mentioned constants as $V^{\prime\prime}(\varphi^{(0)})
\approx (10^{-12} - 10^{-14})\,m_{pl}^2$ and $\Lambda = V(\varphi^{(0)})
\approx m_{pl}^2$, where $m_{pl}$ is the Planck mass. We obtain
a very small number of order of $10^{-12} - 10^{-14}$ that can be
negleted. Thereupon, during the oscillatory phase a selected spectrum
of modes, centered about the integers $j^2 = 2\,k^2$, grow rapidly,
whereas the remaining modes are maintained stable. Suppose that, after
this regime, the universe escapes to inflation. The equation of motion
(\ref{eq1}) assumes its usual form as discussed in several works. As
we know the perturbed modes that cross the horizon are not affected by
microphysical processes, being in this way frozen until they reenter
the horizon after inflation. However, due to the mechanism of parametric
resonance, the modes which cross the horizon for the first time have not
the same amplitude, and are maintained during the kinematical evolution
until they reenter the horizon. At the end, the resulting spectrum of
adiabatic perturbations is not a flat one, in the sense that there are
maxima dictated by Eq. (\ref{max}). The non-flat spectrum of inhomogeneous
fluctuations produced by the above mechanism can explain the hierarchy
of scales (galaxies, clusters, etc) as observed in the actual universe.

The authors are grateful to CNPq and FAPERJ for financial support.


\end{document}